\documentclass[preprint,aps,showpacs,superscriptaddress,nofootinbib]{revtex4}
\usepackage[intlimits]{amsmath}
\usepackage{amssymb}
\usepackage{graphicx}
\usepackage{booktabs}
\usepackage{mathrsfs,slashed}
\usepackage{hyperref}
\usepackage{epsfig}
\usepackage{color}

\newcommand{\bea}{\begin{eqnarray}}
\newcommand{\eea}{\end{eqnarray}}

\begin{document}
\title{Effects of composite pions on the chiral condensate within the 
PNJL model at finite temperature
}

\author{D.~Blaschke}
\email{david.blaschke@ift.uni.wroc.pl}
\affiliation{Institute of Theoretical Physics,
    University of Wroclaw,
    50-204 Wroclaw, Poland}
\affiliation{Laboratory of Theoretical Physics,
    Joint Institute for Nuclear Research,
    141980 Dubna, Russia}
\affiliation{National Research Nuclear University (MEPhI),
    115409 Moscow, Russia}

\author{A.~Dubinin}
\email{dubinin.aleksandr90@gmail.com}
\affiliation{Institute of Theoretical Physics,
    University of Wroclaw,
    50-204 Wroclaw, Poland}
\affiliation{Marian Smoluchowski Institute of Physics,
	Jagellonian University Cracow,
	30-348 Cracow, Poland}

\author{D.~Ebert}
\email{debert@physik.hu-berlin.de}
\affiliation{   Institut f\"ur Physik,
    Humboldt Universit\"at zu Berlin,
    12489 Berlin, Germany}

\author{A.~V.~Friesen}
\email{avfriesen@theor.jinr.ru}
\affiliation{Laboratory of Theoretical Physics,
    Joint Institute for Nuclear Research,
    141980 Dubna, Russia}

\date{\today}
\begin{abstract}
We investigate the effect of composite pions on the behaviour of the chiral condensate 
at finite temperature within the Polyakov-loop improved NJL model.
To this end we treat quark-antiquark correlations in the pion channel (bound states and scattering continuum) within a  Beth-Uhlenbeck approach that uses medium-dependent phase shifts.
A striking medium effect is the Mott transition which occurs when the binding energy vanishes and the discrete pion bound state merges the continuum.
This transition is triggered by the lowering of the continuum edge due to the chiral restoration transition.
This in turn also entails a modification of the Polyakov-loop so that the SU(3) center symmetry gets broken at finite temperature and dynamical quarks (and gluons) appear in the system, taking over the role of the dominant degrees of freedom from the pions.  
At low temperatures our model reproduces the chiral perturbation theory result for the chiral condensate while at high temperatures the PNJL model result is recovered.
The new aspect of the current work is a consistent treatment of the chiral restoration transition region within the Beth-Uhlenbeck approach on the basis of mesonic phase shifts for the treatment of the correlations. 
\end{abstract}

\maketitle


\section{Introduction}

Effective field-theoretic models for low-energy QCD of the Nambu--Jona-Lasinio (NJL) type are widely used to describe the effect of dynamical breaking of the approximate chiral symmetry in the vacuum 
\cite{Ebert:1982pk,Volkov:1982zx,Volkov:1984kq,Ebert:1985kz,Ebert:1994mf}
and its restoration at finite temperatures and chemical potentials \cite{Klevansky:1992qe,Hatsuda:1994pi,Buballa:2003qv}. 
Thereby the chiral condensate and the dynamically generated quark mass gap are considered as synonymous order parameters accessible already on the quark mean-field level of description.
However, it is a well-known problem that the temperature dependence of the melting of the chiral condensate is incorrect. Instead of the behaviour known from the treatment of pionic thermal excitations within chiral perturbation theory which obey Bose statistics, the melting of the chiral condensate with increasing temperature within the mean-field NJL model is due to the fermionic thermal quark excitations. For a discussion see, e.g., Buballa \cite{Oertel:2000jp,Buballa:2003qv} and references therein.   

The alternative of a description of the chiral condensate within a hadron resonance gas model
\cite{Jankowski:2012ms} provides an excellent interpretation of the results from lattice QCD simulations of this quantity which, however, necessarily breaks down at temperatures $T\sim 200$ MeV where the hadron resonance gas model grossly overestimates the number of degrees of freedom and their excitation can no longer be considered as a perturbation. 
Moreover, the underlying assumption of a mixture of ideal gases of hadronic species (statistical model) becomes invalid when the phase space occupation becomes large at temperatures exceeding the pseudocritical temperature $T=154\pm 9$ MeV found in lattice QCD simulations \cite{Bazavov:2011nk}. 
Above this temperature, the effective description with NJL mean-field models, eventually improved by the coupling to a gluon background field with a mean-field pressure modeled by the minimum of the effective potential for the traced Polyakov-loop 
provides an acceptable interpretation of lattice QCD thermodynamics \cite{Ratti:2005jh}.

The question arises whether these two asymptotic situations could be joined in a physically consistent way to a unified description of quark-gluon-hadron thermodynamics and in particular of the chiral condensate. To best of our knowledge such a description does not exist yet.
In order to tackle this problem we would like to follow the scheme of a relativistic 
Beth-Uhlenbeck description of hadronic correlations (bound and scattering states) in quark matter which was advanced by H\"ufner et al. \cite{Hufner:1994ma,Zhuang:1994dw} and recently reformulated within a path integral approach \cite{Wergieluk:2012gd} whereby also diquark correlations were taken into account \cite{Blaschke:2013zaa} as a prerequisite to treat baryons within this approach. In order to mimic quark confinement, the coupling to the Polyakov-loop has been performed in \cite{Wergieluk:2012gd,Yamazaki:2012ux} and the generalization to three quark flavors was done in Refs.~\cite{Yamazaki:2013yua,Dubinin:2016wvt,Torres-Rincon:2017zbr}. 

The important characteristic feature of the generalized Beth-Uhlenbeck approach is that hadronic correlations are treated as bound and scattering states of quarks where the statistical spectral weight for the thermodynamics is given by the phase shift that follows directly from a polar representation of the complex hadronic propagator \cite{Blaschke:2013zaa} as a solution of the corresponding polarization loop integral.
Within the Matsubara formalism, these hadronic phase shifts become dependent on the temperature (and chemical potential) so that also the situation can be described that a bound state merges the continuum of unbound scattering states (Mott effect).  
With this approach, the compositeness aspect of hadrons, in particular also of mesons 
\cite{Blaschke:2017boi}
like the pion as the most important low-energy excitation of the QCD vacuum can be treated in a consistent way, with phase shifts fulfilling the Levinson theorem \cite{Wergieluk:2012gd}.   

In the present work we will for the first time consider the chiral condensate within this approach and, for sake of clarity, will restrict ourselves just to the composite pion as representative of the hadronic degrees of freedom. 
Moreover, we suggest a very schematic model for the continuum phase shift in the form of a step-down function of the energy which represents an "anti-bound state" at the continuum threshold. With this simple model we shall be able to capture the essential features and at the same time allow a semianalytic treatment. In concluding this letter, we shall discuss further developments towards a more realistic description of the thermodynamics of a quark-gluon-hadron system that could be compared to lattice QCD.  

\section{Chiral quark condensate within an effective chiral quark composite meson (CQCM) model}

As a starting point, we use the decomposition of the thermodynamical potential into a 
quark-gluon sector which is treated in the meanfield approximation to the SU(2) PNJL model and a hadronic part which is described by a Mott-hadron resonance gas, here restricted to just the pion channel for simplicity
\bea
\Omega(T ) = \Omega_{PNJL}(T) + \Omega_{\pi}(T).
\eea
The quark-gluon thermodynamic potential at vanishing chemical potential $\mu=0$ is given as
\bea
\Omega_{PNJL}(T) = -4\int\frac{d^3p}{(2\pi)^3}\left\{N_c E({p}) +2T\ln\left[N_\Phi(E(p)) \right]\right\} +\frac{\sigma^2}{4G} +\mathcal{U}(\Phi;T)~,
\label{pPNJL}
\eea
where $N_\Phi(E)=1+3\Phi {\rm e}^{-\beta E}+3\Phi {\rm e}^{-2\beta E}+{\rm e}^{-3\beta E}$ and $E({p})=\sqrt{p^2+m^2}$ with $m=m_{0}+\sigma$ being the constituent quark mass, $m_0=5.5$ MeV the current quark mass and $\sigma$ the dynamically generated quark mass gap. 
For the regularization of the divergent momentum integral over the vacuum energy term we introduce the cutoff $\Lambda=639$ MeV.
According to the standard parametrization scheme of the (P)NJL model (see, e.g., \cite{Grigorian:2006qe})
for the coupling constant we take $G\Lambda^2 = 2.13$, for which the constituent quark mass in vacuum is $m=319$ MeV. The vacuum pion properties ($m_\pi=140 $ MeV, $f_\pi=93$ MeV) and a chiral condensate $\langle \bar{q} q\rangle = -(244~{\rm MeV})^3$ are reproduced. 
{The effective potential was chosen  with the polynomial form and parameters taken from \cite{Ratti:2005jh} with $T_0 = 0.185$ GeV}
\bea\label{effpot}
\frac{\mathcal{U}\left(\Phi;T\right)}{T^4}
&=&-\frac{b_2\left(T\right)}{2} \Phi^2-
\frac{b_3}{3}\Phi^3+
\frac{b_4}{4}\Phi^4, \\ 
b_2\left(T\right)&=&a_0+a_1\left(\frac{T_0}{T}\right)+a_2\left(\frac{T_0}{T}
\right)^2+a_3\left(\frac{T_0}{T}\right)^3~.
\eea
The value of $\Phi$ in thermodynamic equilibrium is obtained from the stationarity condition for the thermodynamic potential
\bea
\frac{\partial \Omega(T)}{\partial \Phi} = 0~,
\eea
which corresponds to a gap equation for this order parameter for deconfinement.
Since we deal with homogeneous systems only, we obtain the pressure as $p(T)= - \Omega(T)$.

The composite meson in our case is the pion and we describe its contribution to the thermodynamics  within the Beth-Uhlenbeck approach \cite{Blaschke:2013zaa,Wergieluk:2012gd}
\bea
\label{ppion}
p_\pi(T) &=&d_\pi \int_{0}^{\infty} \frac{d\omega}{2\pi} \int \frac{d^3q}{(2\pi)^3} 
\delta_\pi (\omega,\vec{q};T) [1+2g(\omega)]~,
\eea
where $g(\omega)=1/[\exp(\beta \omega)-1]$ is the Bose distribution function and 
$\delta_\pi (\omega,\vec{q};T)$ is the quark-antiquark  phase shift in the pion channel.
Instead of using the numerical solution for the Gaussian approximation to the ({P})NJL model as given in \cite{Blaschke:2013zaa}, we suggest in this work the schematic model for the continuum part of the spectrum in the form of an "anti-bound state" at the continuum threshold for 
$\omega =  E_{\rm thr}(q)$.
This model for the phase shift is defined by the ansatz 
\bea
\label{phaseshift}
\delta_\pi (\omega,\vec{q};T)=\pi \left[\Theta(\omega-E_\pi(q))-\Theta(\omega-E_{\rm thr}(q)) \right]
\Theta(T_{\rm Mott}-T)\Theta(|\vec{q}|-\Lambda_\pi)~,
\eea
where $E_\pi(q)=\sqrt{q^2 + m_\pi^2}$ and  $E_{\rm thr}(q)=\sqrt{q^2 + m_{\rm thr}^2}$ with 
$m_{\rm thr}=2m$.
In this ansatz, the $\Theta-$function in the temperature direction makes sure that the density of states stays positive definit and vanishes for $T>T_{\rm Mott}$.
The cutoff $\Lambda_\pi$ may be understood as a consequence of the fact that the nonperturbative interaction in the (P)NJL model gets switched off for momenta above the cutoff $\Lambda$ of the quark sector momentum integral. Thus for high quark momenta chiral symmetry is no longer broken and consequently the Goldstone mode ceases to exist. 
In a recent detailed study of the backreaction effect from the pion and sigma channel on the quark dynamics 
\cite{Kitazawa:2014sga} one observes in Fig.~3 of that work that the pion bound state merges the continuum threshold at a momentum $\sim 600$ MeV which in this case is of the order  $\Lambda$.
With our ansatz of Eq.~(\ref{phaseshift}), there is no correlation in the continuum and therefore the phase shift gets switched off once the binding energy vanishes at the Mott temperature.

After partial integration in Eq.~(\ref{ppion}) over $\omega$ we obtain for the pion pressure\footnote{In the formulation of the (P)NJL model as a separable interaction model \cite{Schmidt:1994di} with a formfactor given by a $\Theta-$function defining the quark momentum cutoff $\Lambda$ the resulting cutoff for the pion momentum integration follows without extra assumptions, see Ref.~\cite{Blaschke:1995gr}} 
\bea
p_\pi(T) &=& -d_\pi \int_{0}^{\infty} d\omega \int
\frac{d^3 q}{(2\pi)^3} 
D_\pi (\omega, q;T) \left[\frac{\omega}{2} +  T \rm{ln}\left(1-e^{-\beta\omega}\right)\right]~.
\eea
The density of states is defined through the  derivative of the pion phase shift (\ref{phaseshift})
\bea
\label{dos}
D_\pi(\omega,q;T) &=& \frac{1}{\pi}\frac{\partial \delta_\pi(\omega,q;T)}{\partial \omega}\nonumber\\
&=&\left[ \delta(\omega - E_\pi(q)) - \delta(\omega - E_{\rm thr}(q)) \right] \Theta(|\vec{q}|-\Lambda_\pi)\Theta(T_{\rm Mott} - T)~,
\eea 
which includes the bound as well as the scattering part  of the spectrum. 


\begin{figure}[!thb]
	\includegraphics[width=0.7\textwidth]{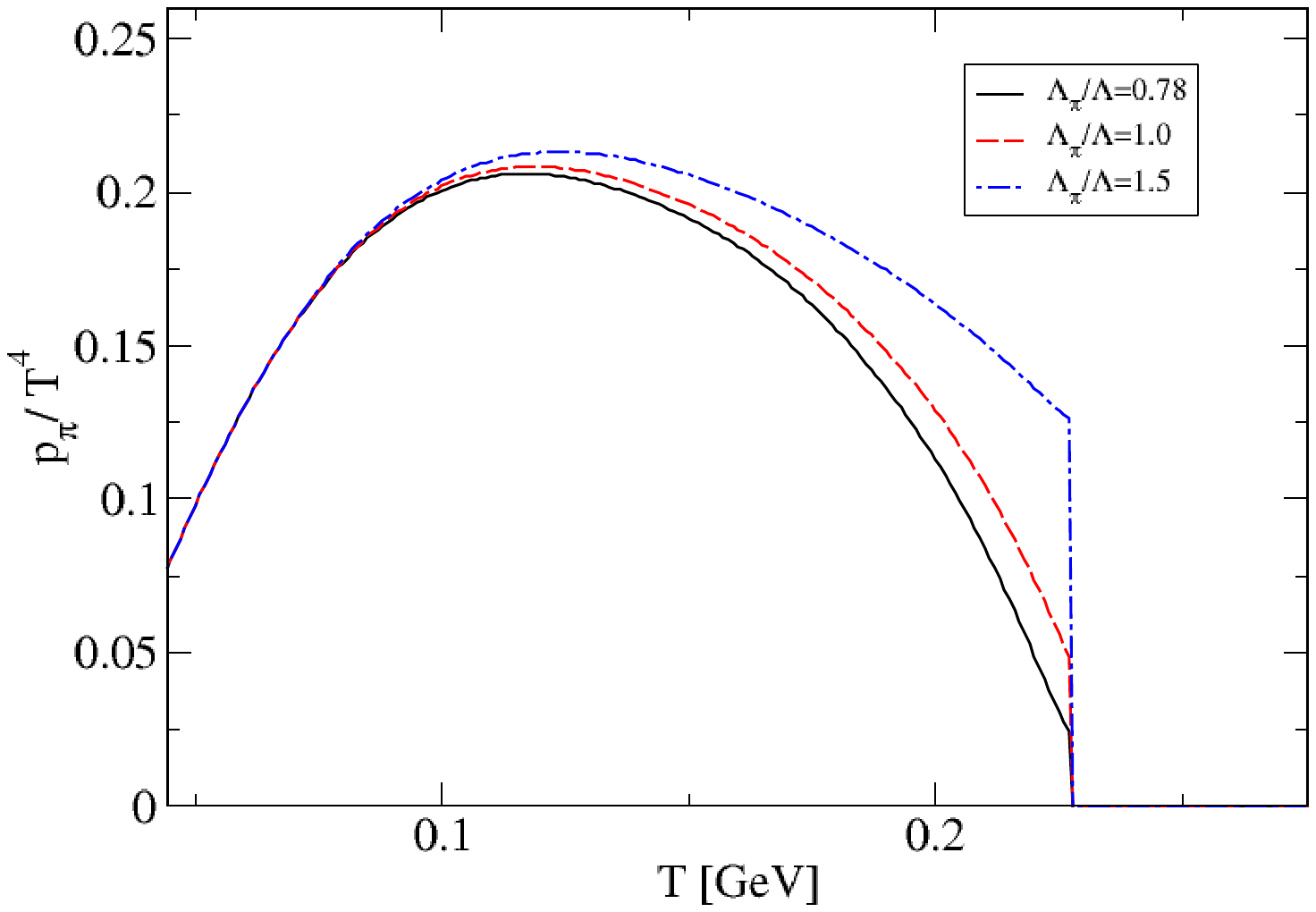}
	\caption{	Pion contribution to the pressure scaled by the factor $T^4$ as a function of the temperature $T$. 		$\Lambda_\pi$ denotes the momentum cutoff for the momentum integral of the pion contribution 		in units of $\Lambda$.
		\label{fig:ppi}
	}
\end{figure}

In this model the pion contribution to the thermodynamics is
\bea
p_\pi(T) &=& -d_\pi \Theta(T_{\rm Mott} - T) \int_{\Lambda_\pi} 
\frac{d^3 q}{(2\pi)^3}
\left\{\frac{E_\pi(q)}{2} - \frac{E_{\rm thr}(q)}{2} \right.\nonumber\\
&&\left. +  T \rm{ln}\left[1-e^{-\beta E_\pi(q)}\right] -  T \rm{ln}\left[1-e^{-\beta E_{\rm thr}(q)}\right]
\right\}~.
\eea

We would like to note that the ansatz (\ref{phaseshift}) for the pion phase shift fulfills the Levinson theorem and that the resulting pressure is a continuous function of the temperature. 
The results for the pion thermodynamics are shown in Fig.~\ref{fig:ppi}.

The remarkable effect of the ansatz for the phase shift fulfilling the Levinson theorem is that the reduction of the pion contribution to the pressure starts already before the Mott temperature is reached, i.e. when the pion is still a true bound state. It is the closeness of the continuum with its negative contribution  to the pressure which makes a qualitatively new effect as compared to the naive hadron resonance gas expectation.

\begin{figure}[!htb]
	\includegraphics[width=0.7\textwidth]{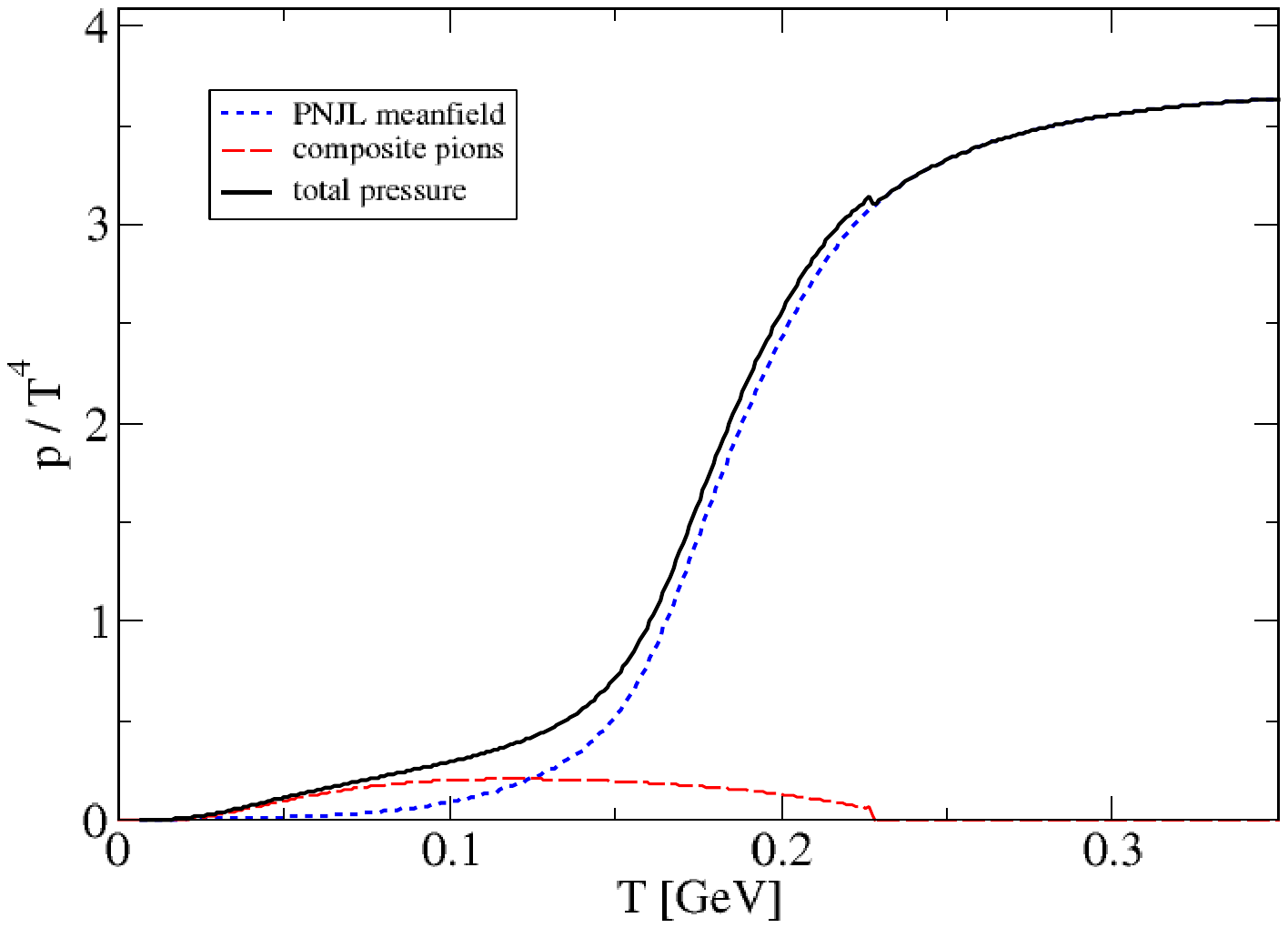}
	\caption{
		Total pressure as a sum of the PNJL meanfield pressure and the pion pressure of Fig.~\ref{fig:ppi} 		scaled by the factor $T^4$ as a function of the temperature $T$.
		\label{fig:ptot}
	}
\end{figure}

The resulting total pressure for the quark-gluon-pion system is shown in Fig.~\ref{fig:ptot}.
We note that for temperatures below $T\sim 100$ MeV the pressure is practically given by that of the pion gas while the PNJL model mean-field pressure that represents the quark-gluon plasma is strongly suppressed
due to the coupling to the traced Polyakov loop $\Phi\sim 0$.
At the deconfinement/chiral restoration transition, the PNJL pressure takes over while the pion gas pressure is strongly reduced due to the influence of the continuum (with negative pressure) and vanishes at the Mott temperature.  

In the following we shall check what is the effect on the chiral condensate.
To this end we recall the standard formula for the chiral quark condensate
\begin{equation}
\langle\bar{q}q\rangle= \frac{\partial\Omega(T)}{\partial m_0}
=\langle \bar{q} q\rangle_{PNJL}+\langle \bar{q} q\rangle_{\pi}~,
\label{qbarq}
\end{equation}
which receives contributions from the PNJL model in mean-field
approximation (see, e.g., Ref.~\cite{Blaschke:2011hm})
\bea
\langle \bar{q} q\rangle_{PNJL} = - 4 N_c \int^\Lambda \frac{dp \ p^2}{2\pi^2} \frac{m}{E({p})} 
\left[1 - 2f_\Phi(E(p))\right]~,
\label{qbarq-q}
\eea
where $f_\Phi(E(p))$ is the Polyakov-loop generalized Fermi distribution function
\cite{Hansen:2006ee,Blaschke:2014zsa}, and from the pion contribution 
\bea
\langle \bar{q} q\rangle_{\pi} &=& \frac{d_\pi}{2}\int^{\Lambda_\pi} \frac{dq\, q^2}{2\pi^2}
\left\{\frac{m_\pi^2}{2m_0 E_\pi(q)}\left[1+2 g( E_\pi(q))\right] \right.\nonumber\\
&& \left. - \frac{2 m_{\rm thr}}{E_{\rm thr}(q)}\frac{\partial m}{\partial m_0} \left[1+2 g( E_{\rm thr}(q))\right]\right\} \Theta(T_{\rm Mott} - T)\, .
\label{qbarq-pi}
\eea
For the derivatives $\partial m_\pi^2 / \partial m_0$ and $\partial m_\pi^2 / \partial \sigma$ in 
Eq.~(\ref{qbarq}) and below in Eq.~(\ref{sigma-gap}) we employ the Gell-Mann--Oakes--Renner (GMOR) 
relation \cite{GellMann:1968rz}, 
\bea
m_\pi^2 f_\pi^2 = - m_0 \langle \bar{q} q \rangle~,
\eea
which for our purpose can be further rewritten by means of the
Goldberger-Treiman relation $f_\pi = m/g_{\pi qq}$ and the relation 
$\langle \bar{q} q\rangle = -(m-m_0)/(2G)$ 
in the more suitable form (see, e.g., \cite{Oertel:2000jp,Blaschke:2017boi,Nikolov:1996jj})
\bea
m_\pi^2 = (m_0/\sigma) g_{\pi qq}^2 /(2G) + O(m_0^2)\, ,
\eea
with $g_{\pi qq}$ being the quark-pion coupling. 
Note further that $\partial m/\partial m_0$ is the chiral susceptibility introduced in \cite{Zhuang:1994dw}.

\begin{figure}[!htb]
	\includegraphics[width=0.7\textwidth]{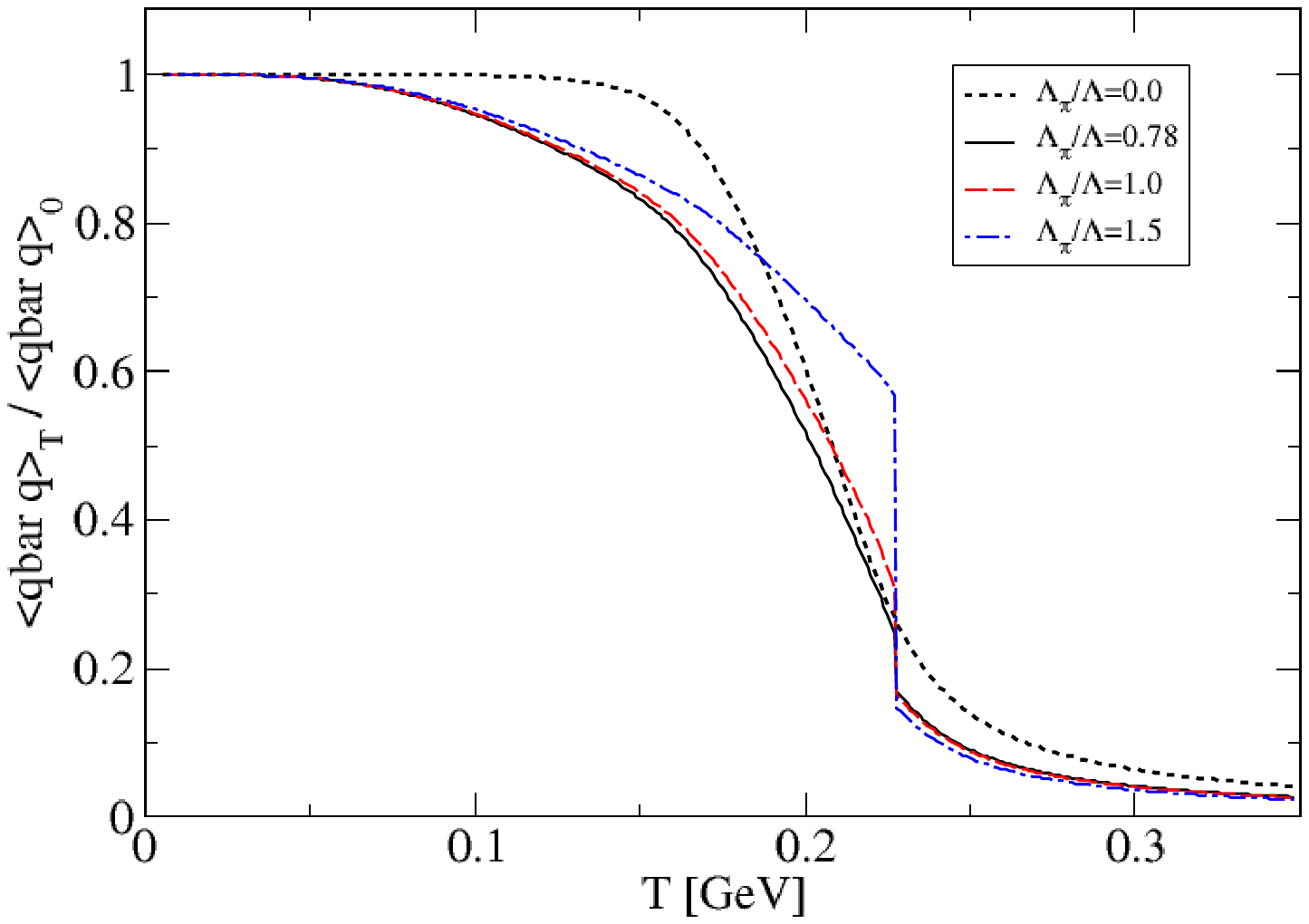}
	\caption{
		Temperature dependence of the chiral condensate in the schematic PNJL quark-gluon-pion model, taking into account the effects from composite pions dominating the medium effects at low temperatures. The different line styles stand for different choices of the pion momentum cutoff $\Lambda_\pi$. Here, only a first step is shown, with a schematic ansatz of a bound state (pion) and an antibound state at the two-quark threshold, mimicking the effect of a continuum and in a very crude manner obeying the in-medium Levinson theorem. The step-like behaviour at the Mott temperature is due to the simplistic ansatz for the continuum phase shift as a step-down function (antibound state) and will become smoothened out as soon as an improved ansatz is used instead, e.g., \cite{Blaschke:2015nma,Blaschke:2016fdh}. 
		\label{fig:cond}
	}
\end{figure}

The results for the temperature dependence of the chiral condensate are shown in Fig.~\ref{fig:cond}. 
One observes the striking effect of the inclusion of the pionic correlations beyond the  PNJL mean-field result (dotted line) for the low-temperature behaviour of the chiral condensate which now exhibits the correct low-temperature limit.
Approaching the Mott temperature from below, the influence of the quark-antiquark cintinuum becomes apparent and its size depends on the chosen value for the cutoff $\Lambda_\pi$ on the pion momentum integral. 
The step-like behaviour at the Mott temperature is well understood as an artefact of the simplistic ansatz for the continuum phase shift as a step function, leading to an abrupt switching off of the pionic correlations above the Mott temperature. 
An improvement of the model ansatz for the continuum phase shift as given, e.g., in Refs. 
\cite{Blaschke:2015nma,Blaschke:2016fdh} will immediately remove this artefact and result in a smooth transition from the region of hadron dominance to quark-gluon dominance in the thermodynamics when passing the Mott temperature. 

\begin{figure}[!htb]
	\includegraphics[width=0.7\textwidth]{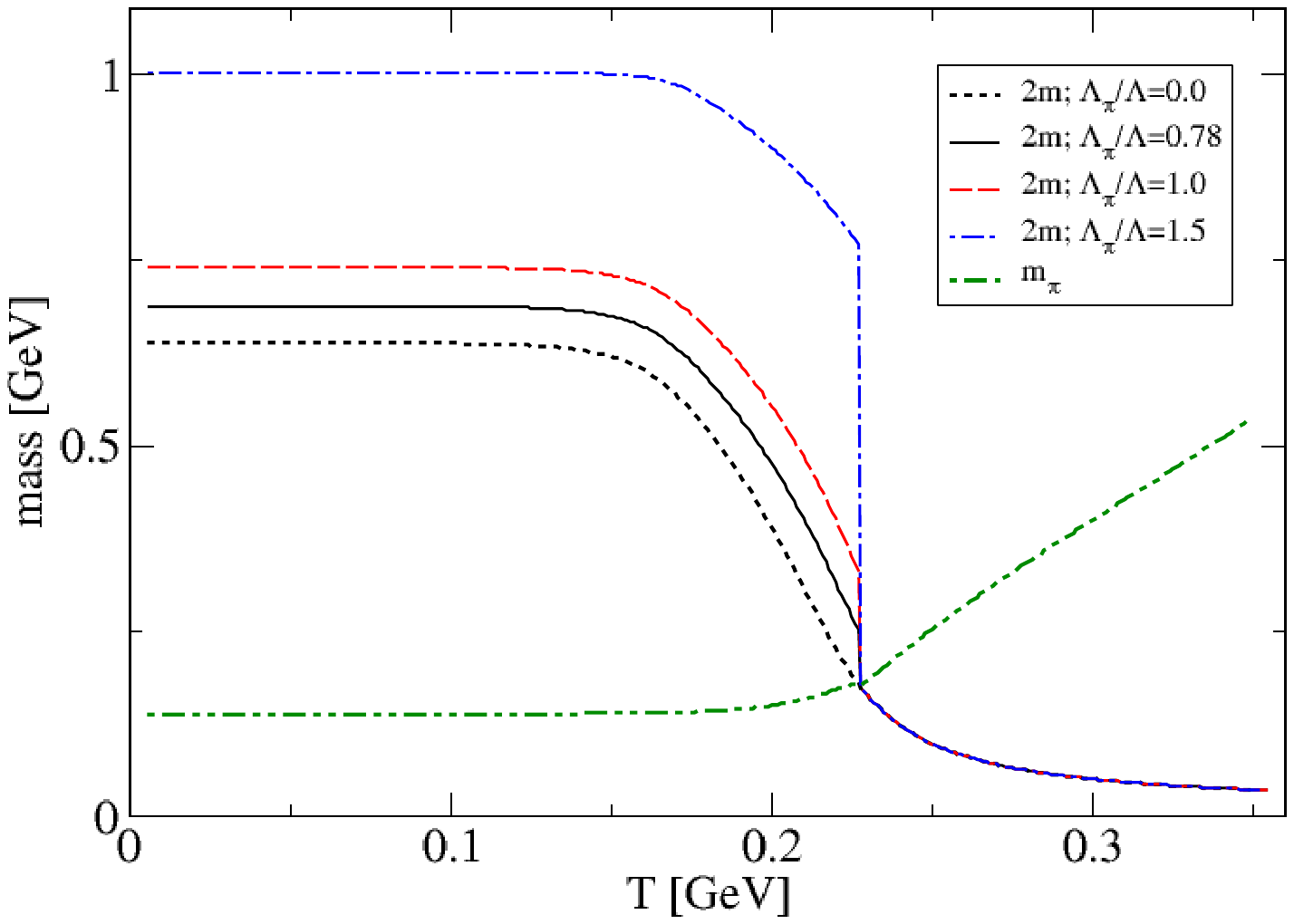}
	\caption{
		Temperature dependence of threshold for the two-quark continuum $m_{\rm thr}=2m$ and the pion mass $m_\pi$ for different values of the cutoff $\Lambda_\pi$ in units of $\Lambda$.
		\label{fig:mass}
	}
\end{figure}

The value of the quark mass gap $\sigma$ one obtains from the solution of the gap equation for $\sigma$ which follows from the condition of stationarity of the thermodynamic potential\footnote{Let us note that in the integrals of Eqs. (\ref{sigma-gap}), (\ref{qbarq-q}) and (\ref{qbarq-pi}) we have placed the regularization cutoff not only at the otherwise divergent vacuum contributions, but also at the thermal ones which in principle do not require regularization. For the present model this is admissible since the thermal contributions from pions are cut off for temperatures $T>T_{\rm Mott}$ where the mean pion momentum would exceed the value of the cutoff. 
	It has been discussed in detail in Refs.~\cite{Florkowski:1993br,Florkowski:1996qb} that both, the vacuum and the matter parts of the meson correlation functions should be regularized in order to maintain important cancellations between them, even though the matter part would not require regularization in order to remove divergences. 
	It is important that no cutoff has been placed at the thermal contributions to the PNJL mean-field pressure in Eq.~(\ref{pPNJL}); otherwise one would not reach the Stefan-Boltzmann limit at high temperatures, see \cite{Zhuang:1994dw}.}
\bea
\frac{\partial \Omega(T)}{\partial \sigma} = 0 &=&\frac{\sigma}{2 G}
- 4 N_c \int^\Lambda \frac{dp \ p^2}{2\pi^2} \frac{m}{E({p})} 
\left[1 - 2f_\Phi(E(p))\right] \nonumber\\
&&- \frac{d_\pi}{2}\int^{\Lambda_\pi} \frac{dq\, q^2}{2\pi^2}\left\{
\frac{m_{\pi}^2}{2\sigma E_{\pi}(q)} \left[1+2 g( E_{\pi}(q))\right]\right.
\nonumber\\
&& \left. +\frac{2 m_{\rm thr}}{E_{\rm thr}(q)} \left[1+2 g( E_{\rm thr}(q))\right]\right\}
\Theta(T_{\rm Mott} - T)
\label{sigma-gap}
\eea
The numerical results for the solution of the gap equation for the quark mass is shown in 
Fig.~\ref{fig:mass}. 
As a new result of this study one observes an additional contribution to the quark mass beyond the PNJL mean-field (dotted line) due to exclusively the continuum contribution to the pion phase shifts, while the pion bound state pole (second term in Eq.~(\ref{sigma-gap})) does practically not contribute. 
The pion mass at finite temperature (dashed line) is obtained as a solution of the finite-temperature Bethe-Salpeter equation with a polarization loop integral defined with quark propagators obtained in the PNJL mean-field approximation. For details see, e.g., Ref.~\cite{Hansen:2006ee} and references therein. 
The pion mass is chirally protected and thus approximately inert against changes in the quark mass gap which results in a negligible contribution from the pion bound state to the quark mass. 
Note that the crossing of the pion mass with the quark-antiquark continuum defines the Mott temperature.

The different role of the pion pole contribution to the chiral condensate and to the quark mass gap can be analytically estimated by the ratio $\sigma/m_0\sim 60$ from comparing the first term in (\ref{qbarq-pi}) with the second in (\ref{sigma-gap}). 
Note that this is only a qualitative argument neglecting a $\sigma-$dependence of the quark-pion coupling $g_{\pi qq}$. The complete numerical evaluation, however, confirms the statement that the second term in Eq.~(\ref{sigma-gap}) is negligible.
The role of the temperature dependent quark-antiquark continuum to the quark thermodynamics was so far overlooked in the literature. Its inclusion is, however, a requirement resulting from the necessity to fulfil the in-medium Levinson theorem for the phase shifts.
We observe that contribution of the continuum to the quark mass gap (given by the third term in (\ref{sigma-gap})) dominates over that from the pion pole while for the chiral condensate the continuum contribution (second term in Eq.~(\ref{qbarq-pi})) is subdominant. 

For the disturbing step-like behaviour of the quark mass at the Mott temperature holds the same discussion as for the similar behaviour observed for the chiral condensate. It is an artefact of the simplistic treatment of the two-quark continuum which will immediately disappear once an improvement of the continuum phase shifts will be considered.    

\section{Conclusions}

We have investigated the effect of composite pions on the chiral condensate 
at finite temperatures within a Beth-Uhlenbeck approach that employs the PNJL model
as an effective approach to deal with interacting quark and gluon degrees of freedom.
For dealing with the quark-antiquark correlations in the pion channel (bound states and scattering continuum) we have introduced a simple model ansatz for medium-dependent phase shifts which embody as a striking medium effect the Mott transition for the pion bound state 
triggered by the lowering of the continuum edge due to the chiral restoration transition.
This ansatz is in accordance with the Levinson theorem generalized to medium-dependent phase shifts.
At low temperatures our model reproduces the chiral perturbation theory result for the chiral condensate while at high temperatures the PNJL model result is recovered.
While the pressure of the quark-gluon-meson system behaves continuous at the Mott transition, a discontinuous behaviour of the pressure derivatives, the chiral condensate and the quark mass gap results. This, however, is well understood as an artefact of the simplistic ansatz for the quark-antiquark scattering continuum and can be systematically improved.

It is easily possible to employ the ansatz (\ref{dos}) for the density of states also for all other hadronic resonances, with appropriate threshold masses for the given hadronic channel.
With such a generalization of the present approach it will be possible to discuss the effect of hadronic excitations on the pseudocritical temperature of the chiral transition which in PNJL models at the mean field level comes out too large, in contradiction with lattice QCD results.
The generalization of the present approach will provide a considerable improvement in understanding the composition and thermodynamics of the quark-gluon-hadron plasma within a unified approach.

\section*{Acknowledgements}
We thank I. Soudi for pointing out a typo in an earlier version of the manuscript.
A.D. and D.E. acknowledge support by the DAAD partnership program between the Humboldt University Berlin and the University of Wroclaw and the hospitality that was extended to them at these Institutions.
A.D. is grateful for support from NCN under grant No. UMO-2016/21/B/ST2/01492.
The work of D.B. and A.F. was supported by the Russian Science Foundation under grant No. 17-12-01427.

\end{document}